\documentclass[twocolumn,prb,amsmath,amssymb,amsfonts,superscriptaddress,floatfix,showpacs]{revtex4}

\usepackage{graphicx}% Include figure files
\usepackage{dcolumn}% Align table columns on decimal point
\usepackage{bm}% bold math

\newcommand{\vct}[1]{\boldsymbol{#1}}

\begin{document}

\title{Adsorption of Cu, Ag, and Au atoms on graphene including van der Waals interactions} 

\author{Martin Amft}
\email{martin.amft@fysik.uu.se}
\affiliation{%
Department of Physics and Astronomy, Uppsala University, Box 516,
S-751 20 Uppsala, Sweden}%
\author{S\'ebastien Leb\`egue}
\affiliation{
Laboratoire de Cristallographie, R\'esonance Magn\'etique et Mod\'elisations (CRM2, UMR CNRS 7036)
Institut Jean Barriol, Nancy Universit\'e
BP 239, Boulevard des Aiguillettes
54506 Vandoeuvre-l\`es-Nancy,France}
\author{Olle Eriksson}
\affiliation{%
Department of Physics and Astronomy, Uppsala University, Box 516,
S-751 20 Uppsala, Sweden}%
\author{Natalia V. Skorodumova}%
\affiliation{%
Department of Physics and Astronomy, Uppsala University, Box 516,
S-751 20 Uppsala, Sweden}%

\date{\today}% 

\begin{abstract}
We performed a systematic density functional study of the adsorption of copper, silver, and gold adatoms on graphene, especially accounting for van der Waals interactions by the vdW-DF and the PBE+D2 methods. 
In particular, we analyze the preferred adsorption site (among top, bridge, and hollow positions) together with the corresponding distortion of the graphene sheet and identify diffusion paths. 
Both vdW schemes show that the coinage metal atoms do bind to the graphene sheet and that in some cases the buckling of the graphene can be significant.
The results for silver are at variance with those obtained with GGA, which gives no binding in this case.
However, we observe some quantitative differences between the vdW-DF and the PBE+D2 methods.
For instance the adsorption energies calculated with the PBE+D2 method are systematically higher than the ones obtained with vdW-DF. 
Moreover, the equilibrium distances computed with PBE+D2 are shorter than those calculated with the vdW-DF method.
\end{abstract}

% 68.43.Fg Adsorbate structure (binding sites, geometry
%71.15.Mb Density functional theory, LDA, GGA etc
%73.22.-f Electronic structure of nanoscale materials: clusters ...

\maketitle

\section{Introduction}
As of today, it is still a challenge for theory to correctly predict the binding of atoms and molecules on coinage metal surfaces.\cite{Brivio:1999p5977}
Recently reported studies, employing a variety of methods, showed the importance of non-local correlations for the correct description of the physisorption of aromatic molecules on copper, silver, and gold surfaces. \cite{Rohlfing:2008p11388,Romaner:2009p11386,Mercurio:2010p11393,Rauls:2010p11378}
Also the reverse setup of coinage metal atoms and clusters on carbon-based materials such as graphite is a long-standing and not completely resolved issue. 
Already Darby et al. [\onlinecite{Darby:1975p11368}] speculated that dispersion forces, i.e. van der Waals forces, significantly contribute to the adsorption energies of gold atoms on graphite.

In recent years graphene, the two-dimensional building block of graphite, has gained much attention in its own right. \cite{Novoselov:2005p11083,Novoselov:2005p11080,Geim:2007p6485,Katsnelson:2007p11097}
The gapless ultrarelativistic energy spectrum of graphene\cite{CastroNeto:2009p10058} as well as effects such as Klein tunneling\cite{klein} are of fundamental interest from a theoretical point of view.
Moreover, graphene shows the Quantum Hall effect at room temperature, \cite{QHE} and its conductive properties correspond to ballistic transport.
Also, graphene offers enormous possibilities for applications in electronics, sensors, biodevices, catalysis, energy storage etc. \cite{CastroNeto:2009p10058,Geim:2009p11096} that often involve adsorbed coinage metal clusters.
For the utilization of these systems a deeper understanding of their stability and electronic properties is needed.
Adatoms and small clusters of coinage metals are known from experiment to be weakly bound and highly mobile on graphite and graphene. \cite{GANZ:1989p10934,Jackson:1994p11241,Francis:1996p10938,Anton:1998p10931,Schaffner:1998p11231,Anton:2000p10930,Yang:2001p11228} 

A wide range of theoretical studies have been performed on small coinage metal clusters on graphite and graphene. \cite{Wang:2003p10933,Jensen:2004p10871,Wang:2004p10870,Varns:2008p10877,Chan:2008p10033,Wu:2009p10942,Amft:2010b}
We are not aware of any experimental study of the adsorption sites or energies of coinage metal adatoms on graphene.
However, there exists, to our knowledge, only one theoretical study that attempted to evaluate the importance of non-local correlation effects on the binding of silver and gold nanoclusters on graphite. \cite{Jalkanen:2007p10951}
The authors of Ref. [\onlinecite{Jalkanen:2007p10951}] used a number of different semi-empirical methods, based on pair potentials, and concluded that the obtained results are in fact strongly dependent on the choice of the method, calling for more studies on these systems. 

To fill this gap, we undertook a systematic investigation of the adsorption of copper, silver, and gold adatoms on graphene, based on density functional theory (DFT) and accounting for van der Waals (vdW) interactions by the vdW-DF method.\cite{Rydberg:2003p10716,Dion:2004p10731,Dion:2005p10754} as well as by the PBE+D2 method\cite{Grimme:06}
In principle, more advanced schemes such as the Random Phase Approximation\cite{Gunnarsson:1976p10967, Langreth1975,Langreth1977} (RPA) or the Quantum Monte-Carlo (QMC) method\cite{Foulkes:2001p337,Spanu:2009p11457} could  also be used to take the van der Waals interaction into account.
Recently, the vdW-DF method has been compared to the RPA for the case of graphite. \cite{Lebegue:2010p11456}
It was found that, although the agreement was not perfect, the vdW-DF appeared as a quantitative method to take dispersive interactions into account.
The same comparison can be made between PBE+D2 and RPA for the same system.
There, the binding energy for graphite was found to be 55 meV/atom with the PBE+D2,\cite{Lebegue:2010} to compare with the 48 meV/atom obtained with the RPA.
Although the PBE+D2 and vdW-DF methods do not reach the accuracy of advanced quantum chemistry methods,  they are at the moment the only computational tools that allows to treat the dispersive interaction in such a large system as a sheet of graphene with a metallic adatom, together with periodic boundary conditions.

\section{Methods}
We calculated the total energies of Cu, Ag, and Au on either of the top (t), bridge (b), or hollow (h) position on a 5$\times$5 graphene sheet, i.e. 50 carbon atoms, using the calculated C-C bond length of 1.42~\AA. 
While the positions of the adatoms were fixed during all calculations, the carbon atoms in the graphene sheet, except those in the rim of the supercell, were free to relax.
The relaxation procedure was stopped, when the Hellmann-Feynman forces on the carbon atoms, that were allowed to relaxed, were smaller than $5 \cdot 10^{-3}$ eV/\AA.

These scalar-relativistic  \textit{ab-initio} DFT calculations were performed using the projector augmented wave (PAW)\cite{Blochl:1994p10844,Kresse:1999p10843} method as implemented in \textsc{vasp}. \cite{Kresse:1996p6093,Kresse:1996p6092}
The exchange-correlation interaction was treated in the generalized gradient approximation (GGA) in the parameterization of Perdew, Burke, and Ernzerhof (PBE) \cite{PERDEW:1996p6520} and, for comparison, also in the local density approximation (LDA) in the parametrization of Perdew and Zunger. \cite{PERDEW:1981p6143}
A cut-off energy of 600 eV was used and a Gaussian smearing with a width of $\sigma$ = 0.05 eV  for the occupation of the electronic levels.
Spin-polarization was taken into account for all the calculations.
The repeated graphene sheets were separated from each other by 20~\AA\, of vacuum.
A Monkhorst-Pack $\Gamma$-centered $5\times5\times1$ k-point mesh (13 k-points in the irreducible wedge of the Brillouin-Zone) was used for the structural relaxations of the carbon atoms.
We also tested a finer k-point mesh, i.e. $16\times16\times1$ k-point mesh (130 k-points), and found the changes in the geometry and the charge density distribution of the systems to be negligible. 

We accounted for the non-local correlation energies by employing two different methods: the van der Waals density functional (vdW-DF) method\cite{Rydberg:2003p10716,Dion:2004p10731,Dion:2005p10754} as implemented in the J\"ulich Non-Local (JuNoLo) code, \cite{Lazic:2010p10753} and the density functional theory plus long-range dispersion correction (DFT+D2) method \cite{Grimme:06} in the implementation of Ref. [\onlinecite{Bucko:2010JChemPhysA}].
From a practical computational point of view, these methods differ significantly.
While vdW-DF is implemented as an \textit{ab-initio} post-processing method in the JuNoLo code, the PBE+D2 method allows to include the vdW interactions in the self-consistency cycle, i.e. during structural relaxations.
In Ref. [\onlinecite{Thonhauser:2007p10699}] it has been shown, for instance, for a set of small molecular systems that the total energies and binding distances obtained with the vdW-DF as a post-processing method and in a self-consistent implementation agree very well.
As an input to the vdW-DF calculations one uses the charge density of the relaxed structure obtained in DFT with PBE.
In PBE+D2, as well as in PBE, we started from the same initial configuration, i.e. a flat graphene sheet with an adatom fixed above one of the adsorption sites.
During the structural relaxation, all the carbon atoms, except for those in the rim, were free to relax.
Note that in PBE+D2 the van der Waals forces were taken into account during the relaxation.

Within the vdW-DF method, the missing vdW interactions are accounted for by replacing the correlation energy from the GGA calculation by a sum of a strictly local $E_{\mathrm{c}}^{\mathrm{LDA}}$ and a non-local correlation $E_{\mathrm{c}}^{\mathrm{nl}}$ energy:

\begin{equation}
E_{0}^{\mathrm{vdW-DF}} = E_{0}^{\mathrm{PBE}} - E_{\mathrm{xc}}^{\mathrm{PBE}} + E_{\mathrm{x}}^{\mathrm{PBE}} + E_{\mathrm{c}}^{\mathrm{LDA}} + E_{\mathrm{c}}^{\mathrm{nl}}
\end{equation}

\noindent where $E_{0}^{\mathrm{PBE}}$ and $E_{\mathrm{xc}}^{\mathrm{PBE}}$ are the total energy and the exchange-correlation energy, respectively, calculated with PBE.
The non-local contribution to the correlation energy is calculated as:

\begin{equation}
E_{c}^{\mathrm{nl}} = \frac{1}{2} \int d^{3} r \, d^{3} r' n(\vec{r}) \phi(\vec{r},\vec{r}\,')n(\vec{r}\,'),
\end{equation}

\noindent with $n(\vec{r})$ being the charge density.
The kernel $\phi(\vec{r},\vec{r}\,')$ depends on the distance $(\vec{r} - \vec{r}\,' )$ and the electron densities $n$ in the vicinity $\vec{r}$ and $\vec{r}\,'$.

In the PBE+D2 method, on the other hand, the vdW interactions are described by a pair-wise correction ($E_{\text{disp}}$), optimized for some popular DFT functionals, and added to the self-consistent Kohn Sham energy ($E_{\text{KS-DFT}}$) such that:

\begin{equation}\label{eq_energy}
E_{\text{PBE+D2}} = E_{\text{KS-DFT}} + E_{\text{disp}}.
\end{equation}

\noindent with

\begin{equation}\label{e4}
   E_{\text{disp}} = -\frac{s_6}{2} \sum_{i=1}^{N_{at}} \sum_{j=1}^{N_{at}}    
   \sum_{\vct{L}}{}^\prime
   \frac{C_6^{ij}}
        {|\vct{r}^{i,0}-\vct{r}^{j,\vct{L}}|^6} 
   f(|\vct{r}^{i,0}-\vct{r}^{j,\vct{L}}|),
\end{equation}

\noindent where the summations are over all the atoms and all the translations of the unit cell. 
The prime indicates that $i\not=j$ for L=0, ensuring that there is no double counting. 
The scaling factor $s_6$ is dependent on the Kohn-Sham functional.
Its value is 0.75 for the PBE functional, which was used here.
$C_6^{ij}$ is the vdW coefficient for an atom pair ($ij$) and $\vct{r}^{i,\vct{L}}$ is the position vector of atom $i$ after performing $\vct{L}$ translations of the unit cell along lattice vectors. 
Also, $f(r)$ is a damping function which cancels the contribution from $E_{\text{disp}}$ for distances corresponding to standard (covalent or ionic) bonds.
In practice, the summations over all the atoms are replaced by summations up to a suitably chosen radius.
All the pair interactions up to a radius of 12 \AA~were included in the calculation of $E_{\text{disp}}$.
Moreover, since gold was not included in the list of the elements provided in the original DFT+D2 article~\cite{Grimme:06}, we have used a value of 40.62 Jnm$^6$/mol for the C$_6$ coefficient and of 1.772 \AA~for the vdW radius of Au.\cite{Grimme-private}
Note that for the DFT part ($E_{\text{KS-DFT}}$) of the calculation, we used the same scalar-relativistic DFT method and PBE functional as described above.

The adsorption energies $E_{\mathrm{ads}}$ of a metal atom M on a graphene sheet G are calculated as:

\begin{eqnarray*}
E_{\mathrm{ads}}^{\mathrm{PBE+D2}}\!&=&\!E_{0}^{\mathrm{PBE+D2}} \mathrm{[M/G]}\!-\!E_{0}^{\mathrm{PBE+D2}} \mathrm{[G]}\!-\!E_{0}^{\mathrm{PBE}} \mathrm{[M]}\\
  \\
E_{\mathrm{ads}}^{\mathrm{vdW-DF}}\!&=&\!E_{0}^{\mathrm{vdW-DF}} \mathrm{[M/G]}\!-\!E_{0}^{\mathrm{vdW-DF}} \mathrm{[G]}\!-\!E_{0}^{\mathrm{vdW-DF}} \mathrm{[M]} \\
\\
E_{\mathrm{ads}}^{\mathrm{LDA,PBE}}\!&=&\!E_{0}^{\mathrm{LDA,PBE}} \mathrm{[M/G]}\!-\!E_{0}^{\mathrm{LDA,PBE}} \mathrm{[G]}\!-\!E_{0}^{\mathrm{LDA,PBE}} \mathrm{[M]}
\end{eqnarray*}

\noindent where the E$_{0}$ are the ground state energies of the adatom on graphene [M/G], graphene [G] and metal [M] alone, calculated with the marked method.
The $E_{\mathrm{ads}}$ are negative when the adsorption is exothermic.

\section{Results}

In this section, we present our results concerning the adsorption energies, preferred sites, and equilibrium distances for Cu, Ag, and Au adatoms on graphene.
In order to reach a qualitative conclusion, we used two different schemes to take the dispersive interaction into account.
Namely, the vdW-DF and the PBE+D2 methods were employed, see the previous section for details. 
For the sake of comparison, we also report our results obtained with a semi-local (GGA) and a local (LDA) approximation, although none of these methods are able to describe the non-local  correlation effects needed to tackle the present problem  correctly.

\begin{figure}[hbt]
\includegraphics[width=0.9\linewidth]{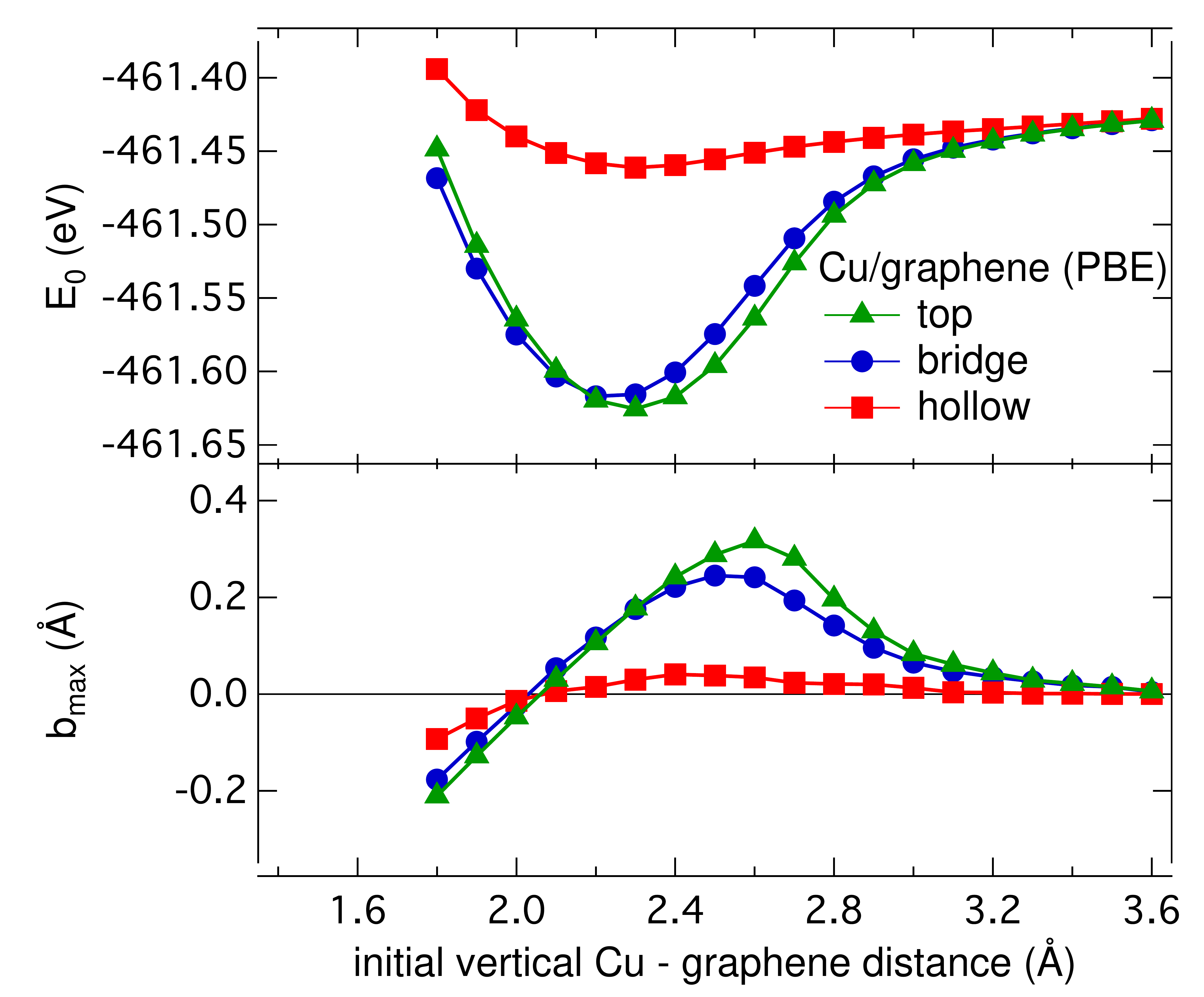}  
 \caption{(Color online) Upper panel: total energy vs. initial vertical Cu  - graphene distance for three binding sites, calculated in PBE. Lower panel: vertical distortion of the carbon atoms closest to the Cu adsorption site.}
\label{fgr:CuPBE}
\end{figure}

\begin{figure}[hbt]
\includegraphics[width=0.9\linewidth]{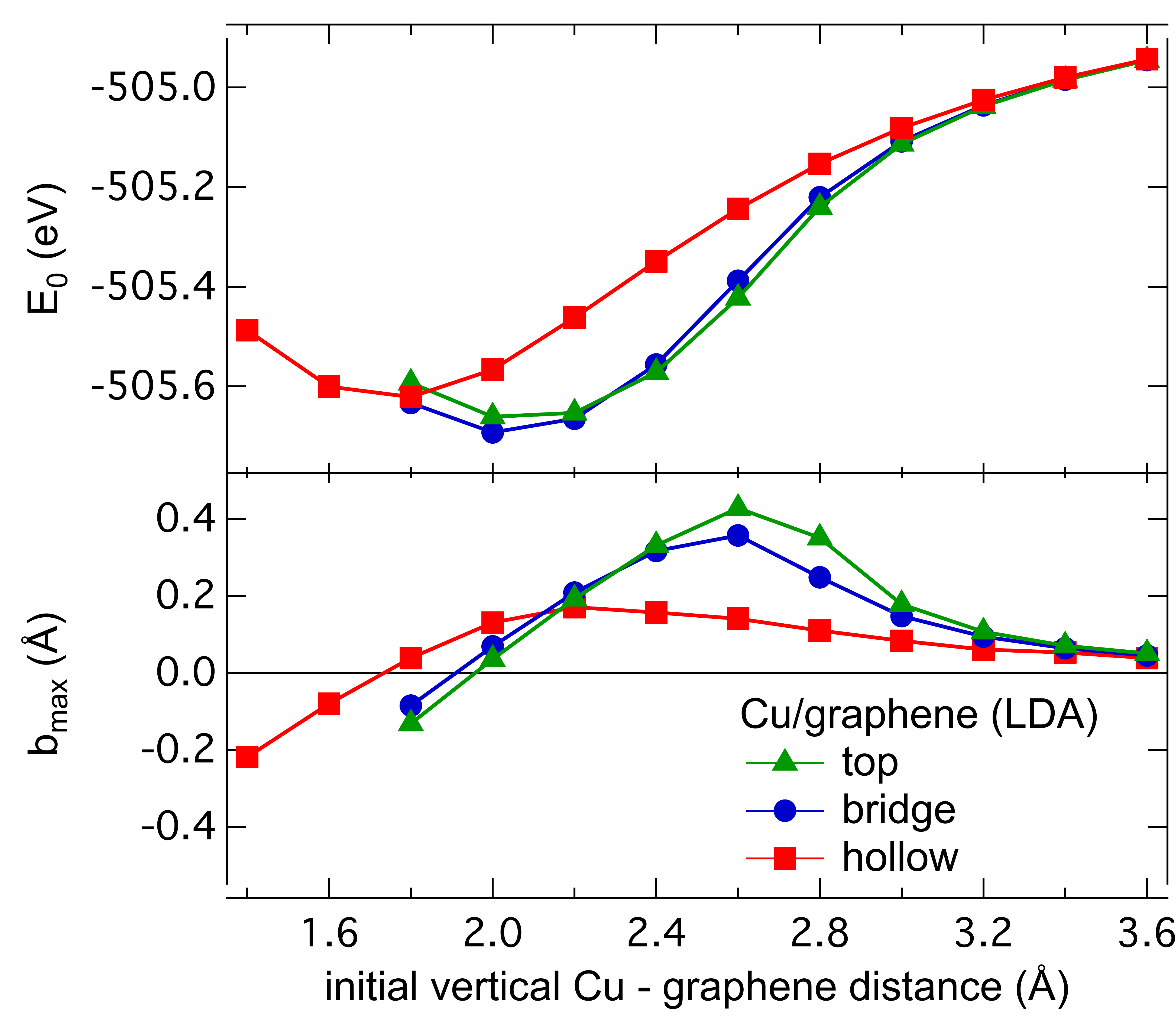}  
 \caption{(Color online) Upper panel: total energy   vs. initial vertical Cu  - graphene distance for three binding sites, calculated in LDA. Lower panel: vertical distortion    of the carbon atoms closest to the Cu adsorption site.}
\label{fgr:CuLDA}
\end{figure}

\begin{figure}[hbt]
\includegraphics[width=0.9\linewidth]{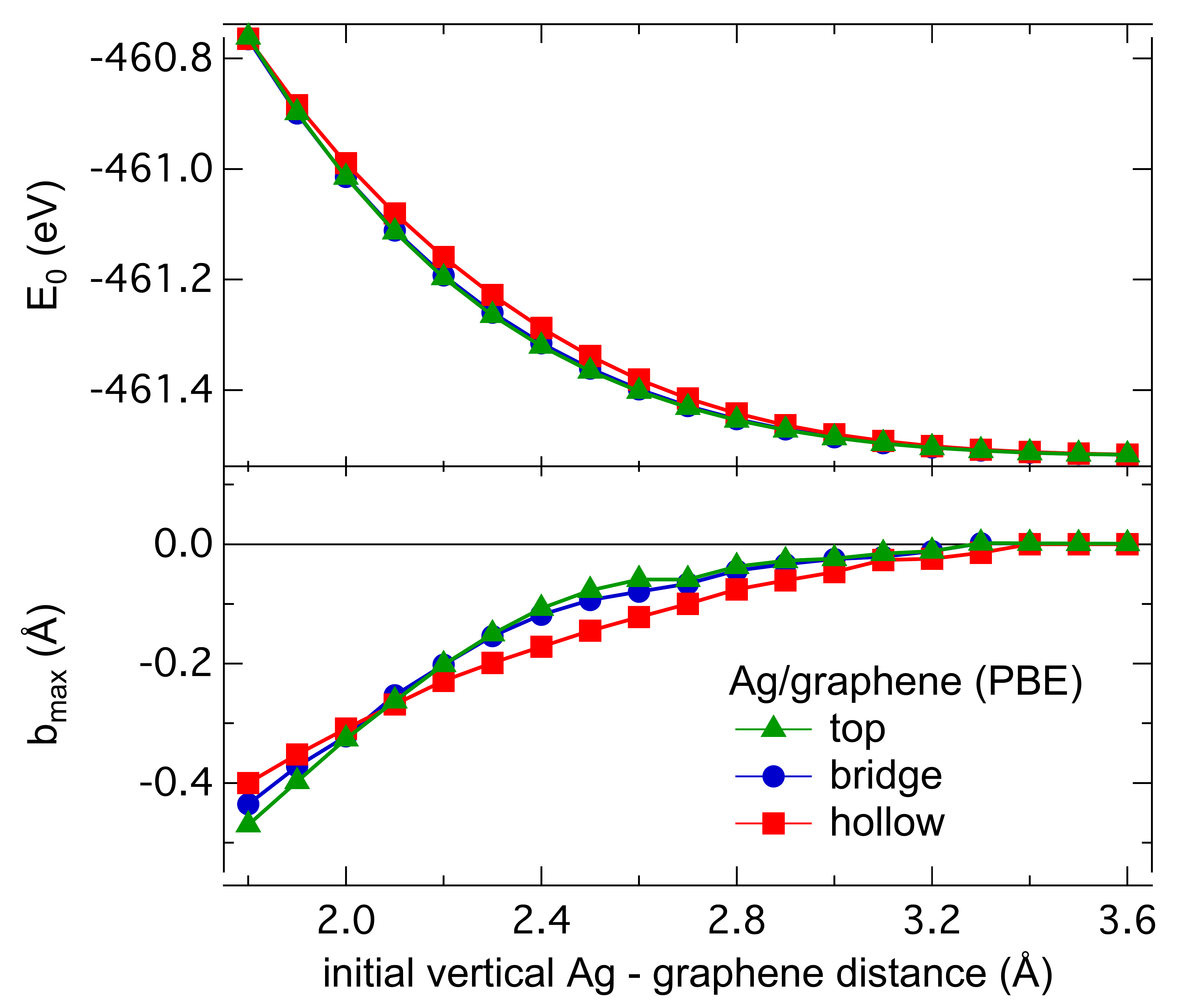}  
 \caption{(Color online) (Color online) Upper panel: total energy   vs. initial vertical Ag  - graphene distance for three binding sites, calculated in PBE. Lower panel: vertical distortion    of the carbon atoms closest to the Ag adsorption site.}
\label{fgr:AgPBE}
\end{figure}

\begin{figure}[hbt]
\includegraphics[width=0.9\linewidth]{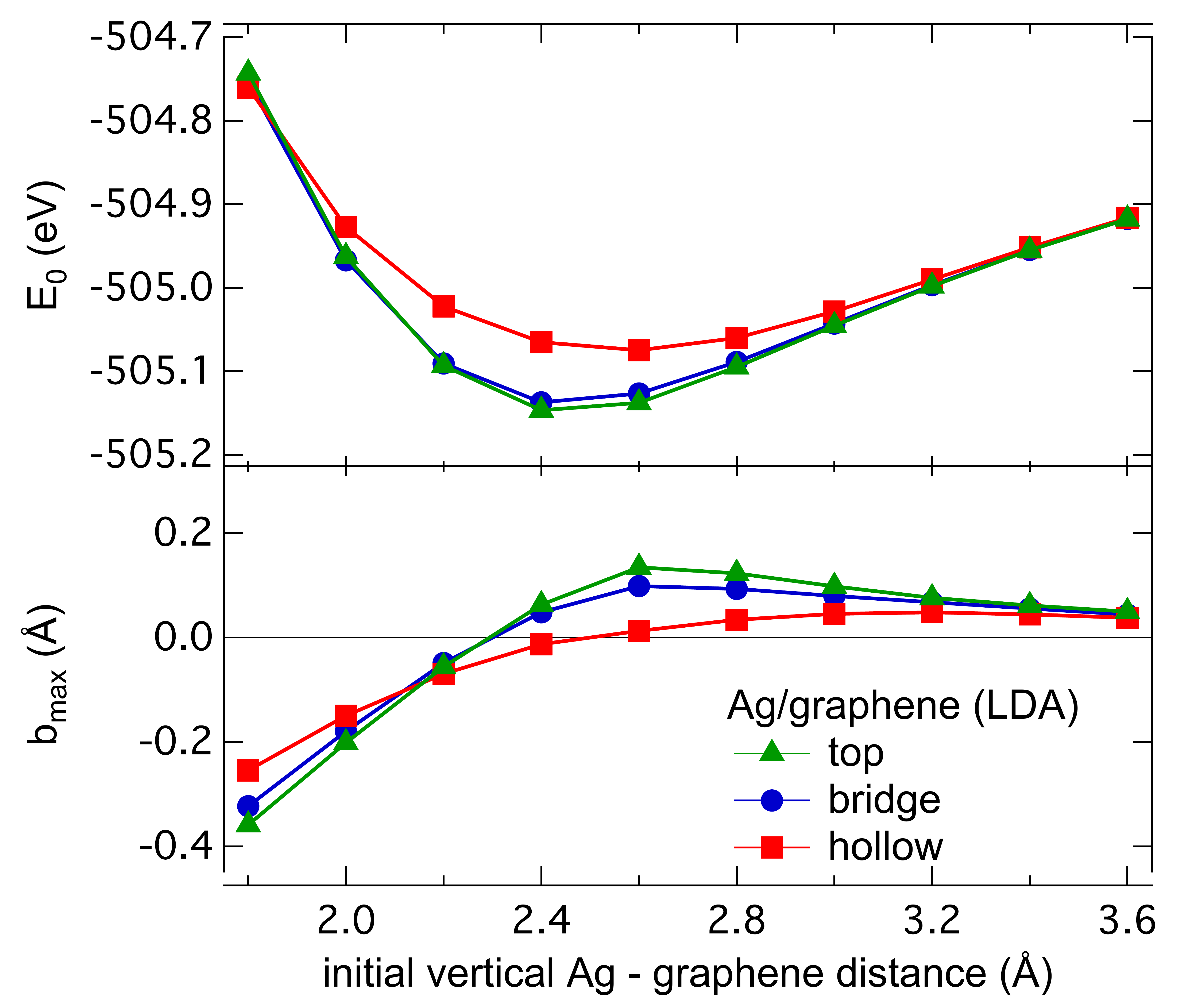}  
 \caption{(Color online) (Color online) Upper panel: total energy   vs. initial vertical Ag  - graphene distance for three binding sites, calculated in LDA. Lower panel: vertical distortion    of the carbon atoms closest to the Ag adsorption site.}
\label{fgr:AgLDA}
\end{figure}

\begin{figure}[hbt]
\includegraphics[width=0.9\linewidth]{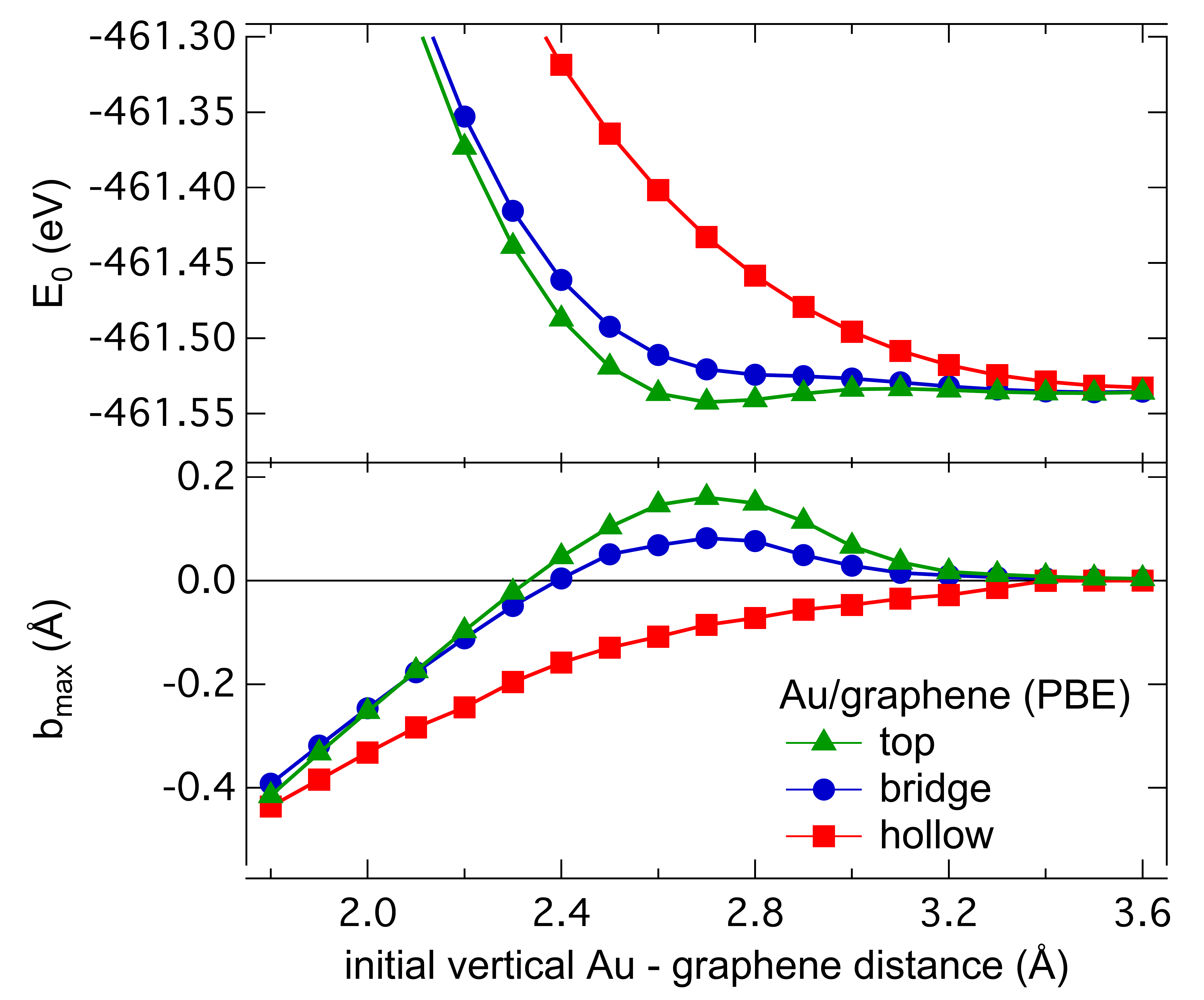}  
 \caption{(Color online) Upper panel: total energy   vs. initial vertical Au  - graphene distance for three binding sites, calculated in PBE. Lower panel: vertical distortion    of the carbon atoms closest to the Au adsorption site.}
\label{fgr:AuPBE}
\end{figure}

\begin{figure}[hbt]
\includegraphics[width=0.9\linewidth]{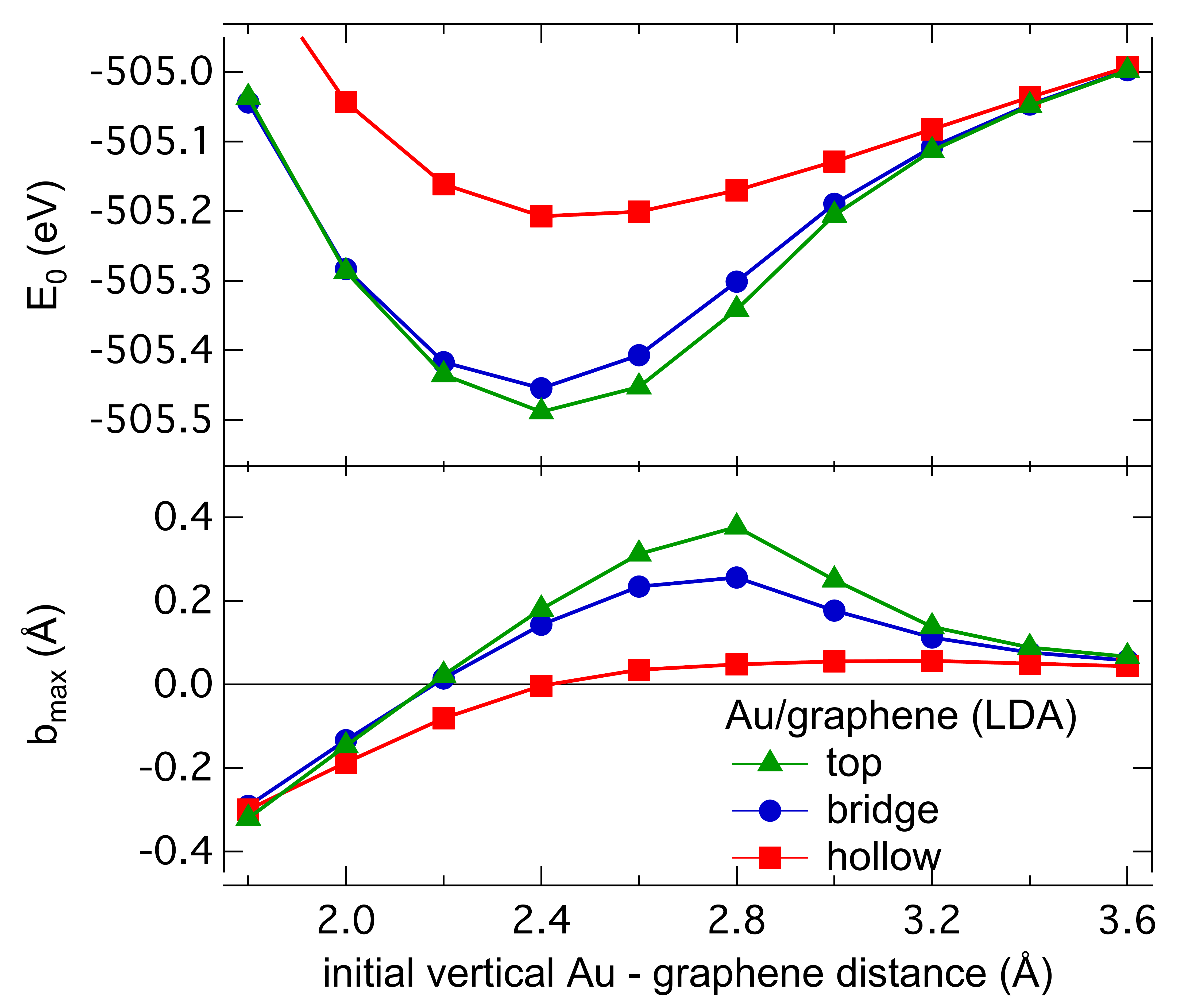}  
 \caption{(Color online) Upper panel: total energy   vs. initial vertical Au  - graphene distance for three binding sites, calculated in LDA. Lower panel: vertical distortion    of the carbon atoms closest to the Au adsorption site.}
\label{fgr:AuLDA}
\end{figure}

The total energy curves, where vdW interactions were taken into account, are shown in Figs \ref{fgr:CuD2} - \ref{fgr:AuvdW}.
In the Appendix, we also show the the PBE and LDA total energy curves, Figs \ref{fgr:CuPBE} - \ref{fgr:AuLDA}, for completion.
In all of the Figs \ref{fgr:CuD2} - \ref{fgr:AuLDA} the upper panels show the total energy $E_{0}$ as a function of the initial vertical metal atom - graphene distance.
The total energies were calculated for three adsorption sites of the metal atom, i.e. top (t), bridge (b), and hollow (h) position, respectively.
The lower panels in Figs \ref{fgr:CuD2} - \ref{fgr:AuLDA} show the maximal vertical distortion ($b_{\mathrm{max}}$) of the carbon atoms in the first coordination shell of the adsorption sites.
Positive values of $b_{\mathrm{max}}$ mean the graphene sheet buckles towards the adatom.

The calculated adsorption energies $E_{\mathrm{ads}}$ and their corresponding vertical equilibrium distances $h_{\mathrm{equ}}$ are compiled in Tables \ref{tableCu}, \ref{tableAg}, and \ref{tableAu}.
The equilibrium distances are the difference between the initial vertical metal atom - graphene distance $h_{\mathrm{init}}$ and the vertical buckling: $h_{\mathrm{equ}} = h_{\mathrm{init}} - b_{\mathrm{max}}$.
Tables \ref{tableCudiff}, \ref{tableAgdiff}, and \ref{tableAudiff} show the differences between the total energies of adsorption sites, e.g.  $\Delta E_{0}^{t-b}$ between the top and bridge position.
These $\Delta E_{0}$ give an indication about the preferential diffusion paths of the adatoms on graphene.

Note that all the systems have a total spin moment of $1\mu_{B}$ due to the single valence electron of the coinage metal atoms. 
 This value of the magnetic moment is essentially preserved even for the shortest distance, showing that no significant charge transfer
  takes place between the adatom and the graphene sheet.

\subsection{Cu on graphene}

\begin{figure}[hbt]
\includegraphics[width=0.95\linewidth]{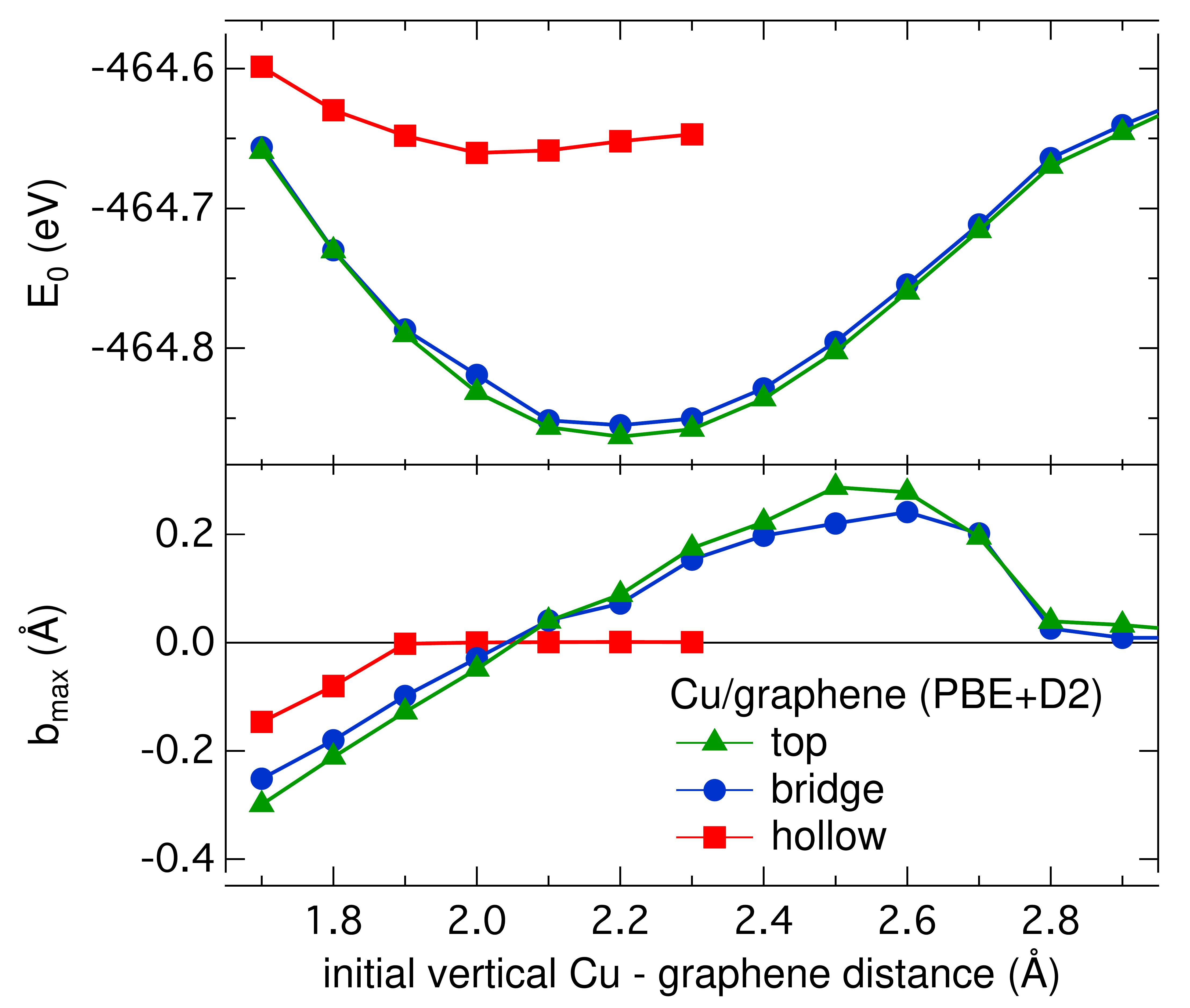}  
 \caption{(Color online) Upper panel: total energy ($E_{0}$) as a function of the initial vertical Cu adatom - graphene sheet distance $h$ for three binding sites, calculated in PBE+D2. Lower panel: vertical distortion ($b_{\mathrm{max}}$) of the carbon atoms in the first coordination shell of the Cu adsorption site. Positive values of $b_{\mathrm{max}}$ mean the carbon atoms buckle towards the adsorbed adatom.}
\label{fgr:CuD2}
\end{figure}

\begin{figure}[hbt]
\includegraphics[width=0.95\linewidth]{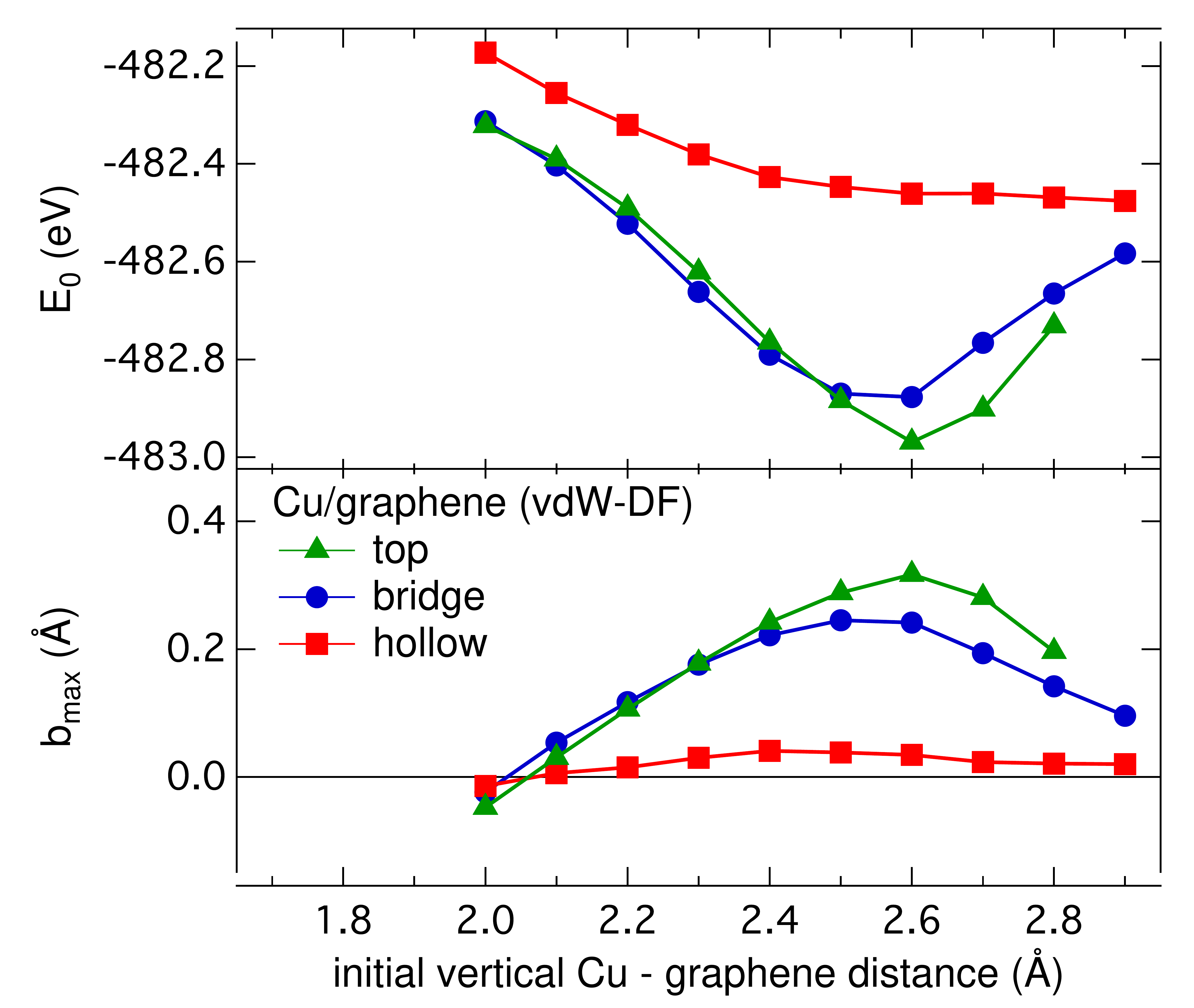}  
 \caption{(Color online) Upper panel: total energy vs. initial vertical Cu  - graphene distance for three binding sites, calculated in vdW-DF. Lower panel: vertical distortion ($b_{\mathrm{max}}$) of the carbon atoms closest to the Cu adsorption site.}
\label{fgr:CuvdW}
\end{figure}

We begin by presenting our results for copper on graphene.
The total energy curves of the PBE+D2 and the vdW-DF calculations for this system are shown in Figs \ref{fgr:CuD2} and \ref{fgr:CuvdW} (upper panels).
For both methods, the top position is most preferable.
Especially in the PBE+D2 results, the top position competes within a small but noticeable energy  difference with the bridge configuration, while the hollow position is definitively not favorable in any of the two methods. 
The PBE (Fig. \ref{fgr:CuPBE} in the appendix) approximation correctly predicts the top site as being the ground state, but all the obtained  minima are comparatively shallow, as one also can see from the adsorption energies in Table \ref{tableCu}.
The ordering of the adsorption sites obtained with the LDA (Fig. \ref{fgr:CuLDA} in the appendix) are in contrast to the aforementioned vdW and PBE results.
In LDA, the three sites are in a competitive energy range, with the bridge position being the preferred one. 

\extrarowheight3pt
\begin{table}[htdp]
\caption{Adsorption energies (eV) of Cu on graphene, calculated by four different approximations to the exchange-correlation functional, initial vertical Cu-graphene distance $h$ and vertical equilibrium distances $h_{\mathrm{equ}}$ (\AA), at three binding sites, top (t), bridge (b), and hollow (h).}
\begin{center}
\begin{tabular}{|c|cc|cc|cc|}
\hline
& \multicolumn{2}{c|} {top} & \multicolumn{2}{c|} {bridge} & \multicolumn{2}{c|} {hollow}\\
& $E_{\mathrm{ads}}^{t}$ & $h_{\mathrm{equ}}^{t}$ & $E_{\mathrm{ads}}^{b}$ &  $h_{\mathrm{equ}}^{b}$ & $E_{\mathrm{ads}}^{h}$ & $h_{\mathrm{equ}}^{h}$  \\ \hline
PBE+D2& \quad -0.909 & 2.11 & \quad -0.901 & 2.13 & \quad -0.706 & 2.00\\ 
vdW-DF & \quad -0.684 & 2.28 & \quad -0.592 & 2.36 & \quad -0.176 & 2.57 \\ 
PBE& \quad -0.228 & 2.12 & \quad -0.219 & 2.08 & \quad -0.063 & 2.27 \\  
LDA & \quad -0.842 & 1.96 & \quad -0.874 & 1.93 & \quad -0.802 & 1.76\\ 
\hline
\end{tabular}
\end{center}
\label{tableCu}
\end{table}%

Table \ref{tableCu} summarizes the adsorption energies $E_{\mathrm{ads}}$ and corresponding  vertical equilibrium distances $h_{\mathrm{equ}}$ for Cu and graphene.
While PBE+D2 gives $E_{\mathrm{ads}}$ of $-0.706$ to $-0.909$ eV, PBE predicts a much weaker binding, ranging from $-0.063$ to $-0.228$ eV, only.
Also, within the resolution of 0.1\,\AA\, PBE and PBE+D2 give similar equilibrium distances for the top and the bridge position of Cu, but the equilibrium distance at the hollow site is 12\%\, smaller in PBE+D2 than in PBE.
The adsorption energies in vdW-DF range from $-0.176$ to $-0.684$ eV, thereby lying in between those of PBE and PBE+D2.
Note that this method gives only a very shallow local minimum at the hollow site, see Fig. \ref{fgr:CuvdW}.
Compared to PBE and PBE+D2, the equilibrium distances in vdW-DF are shifted towards larger values, exceeding those of PBE+D2 by up to 29\%.
Incidentally, LDA gives binding energies quite close to PBE+D2 (within 0.1 eV), but up to 12\%\, shorter equilibrium distances.

In the cases of PBE+D2 and vdW-DF the deformation of the graphene sheet is always positive around the equilibrium distance, i.e. the carbon atoms and the copper atom attract each other, see lower panels in Figs \ref{fgr:CuD2} and \ref{fgr:CuvdW}.
For smaller distances, the buckling becomes negative, which is obviously a steric effect. 
Also, the buckling is always larger for the top configuration, which is an other picture of the increased stabilization of this site with respect to the hollow and bridge ones. 
The vertical distortions of the graphene sheet obtained from the PBE and the PBE+D2 structural relaxations differ by less  than 10\% only.

The binding order in from the LDA calculations is different, see Fig. \ref{fgr:CuLDA}.
The maximum buckling is obtained for a top configuration, although the bridge position is the actual ground state in LDA.
Hence, the energy minimum and buckling do not seem to be correlated in this approximation.

\begin{table}[htdp]
\caption{Identifying the diffusion path of Cu on graphene from the difference in the total energies $\Delta E_{0}$ (meV) at the three adsorption sites. }
\begin{center}
\begin{tabular}{|c|c|c|c|}
\hline
& top to bridge & top to hollow & bridge to hollow \\ 
& $\Delta E_{0}^{t-b}$ & $\Delta E_{0}^{t-h}$ & $\Delta E_{0}^{b-h}$ \\ \hline  
PBE+D2 & 8 & 203 & 195  \\ 
vdW-DF & 92 & 508 & 416  \\ 
PBE & 9 & 164 & 156   \\ 
LDA & 31 & 40 & 71 \\  \hline
\end{tabular}
\end{center}
\label{tableCudiff}
\end{table}%

From the differences of the total energies at the different adsorption sites, $\Delta E_{0}$ in Table \ref{tableCudiff}, one can obtain some insight into the diffusion paths of the coinage metal adatoms  on graphene, since it has been shown that $\Delta E_{0}$ [top - bridge] is the height of the diffusion barrier along the carbon-carbon bonds. \cite{Amft:2010b}
Although the energy differences vary significantly for $\Delta E_{0}^{t-b}$ in the case of copper, the overall trends are identical in PBE+D2 and vdW-DF: diffusion will take place along the carbon-carbon bonds, since a path over the hollow site is energetically too expensive.

\subsection{Ag on graphene}
\begin{figure}[hbt]
\includegraphics[width=0.95\linewidth]{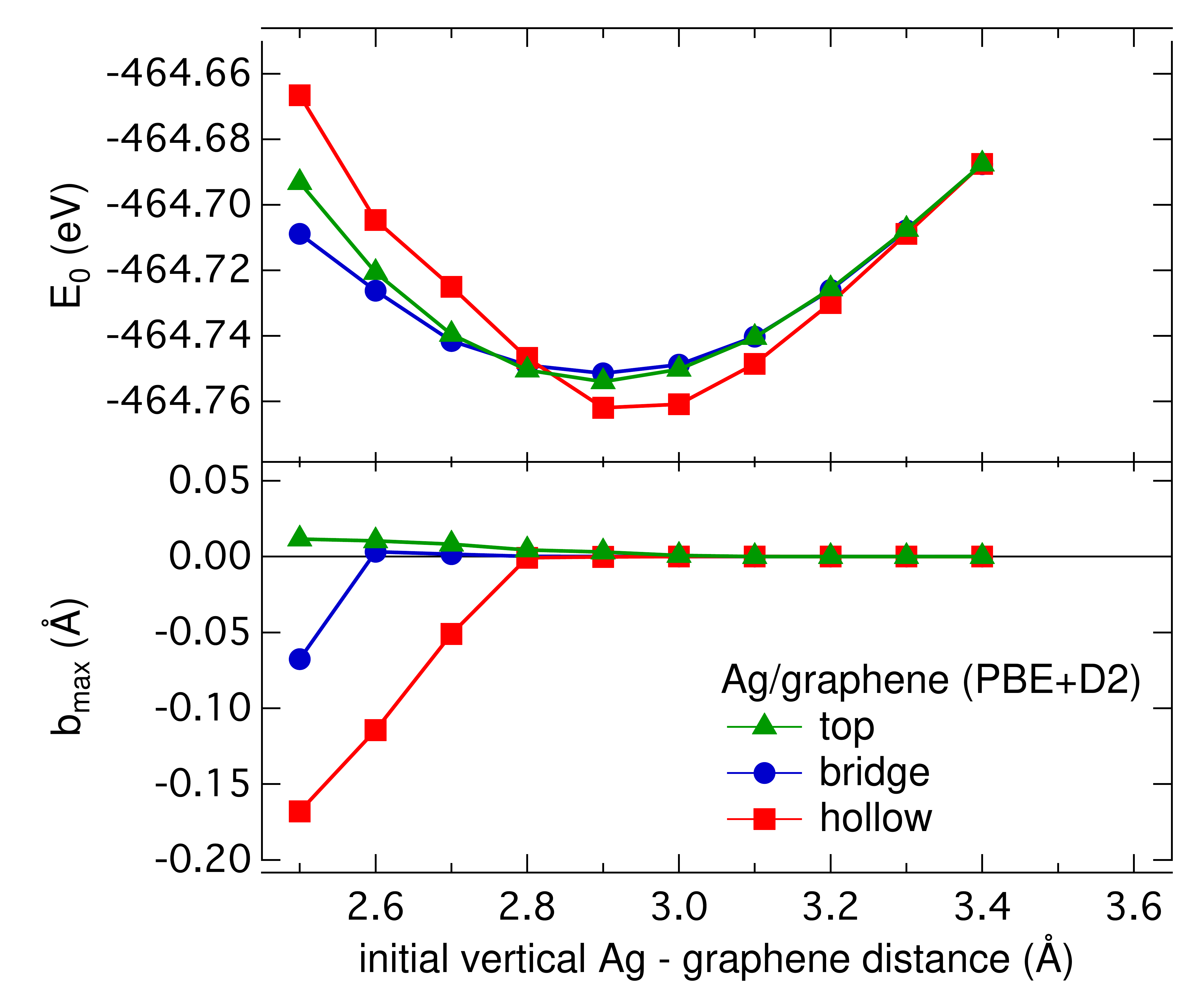}  
 \caption{(Color online) Upper panel: total energy as a function of the initial vertical Ag adatom - graphene sheet distance $h$ for three binding sites, calculated in PBE+D2. Lower panel: vertical distortion ($b_{\mathrm{max}}$) of the carbon atoms in the first coordination shell of the Ag adsorption site. Positive values of $b_{\mathrm{max}}$ mean the carbon atoms buckle towards the adsorbed adatom.}
\label{fgr:AgD2}
\end{figure}

\begin{figure}[hbt]
\includegraphics[width=0.95\linewidth]{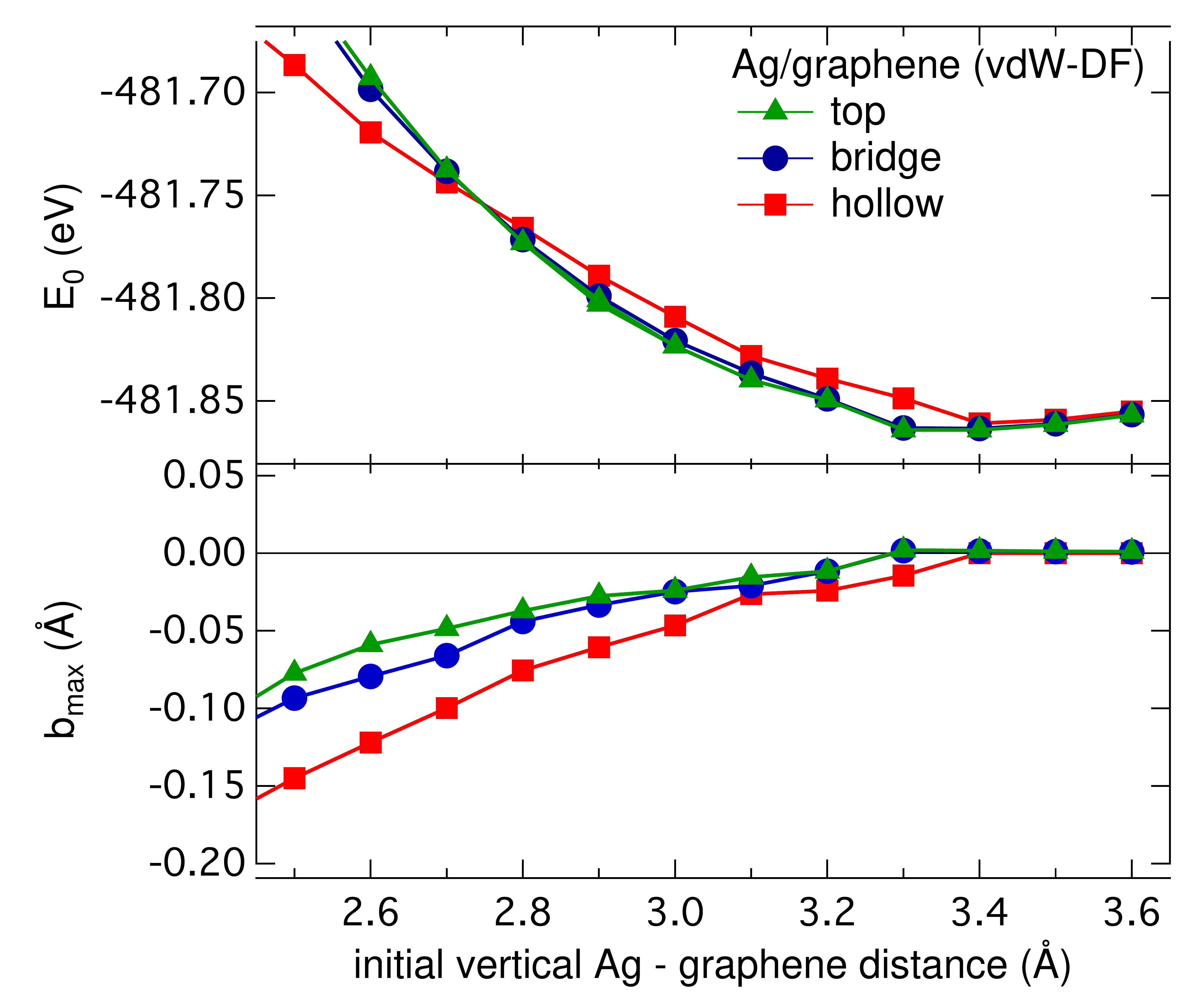}  
 \caption{(Color online) Upper panel: total energy   vs. initial vertical Ag  - graphene distance for three binding sites, calculated in vdW-DF. Lower panel: vertical distortion ($b_{\mathrm{max}}$) of the carbon atoms closest to the Ag adsorption site.}
\label{fgr:AgvdW}
\end{figure}

The silver adatom on graphene presents quite a different picture than copper and gold do.
While PBE+D2 (Fig. \ref{fgr:AgD2}) and vdW-DF (Fig. \ref{fgr:AgvdW}) clearly show binding at large equilibrium distances, i.e. 2.9\,\AA\, and 3.3\,\AA, respectively, pure GGA fails completely to predict any binding, see Fig. \ref{fgr:AgPBE}.
Note that we repeated the PBE calculations with other GGA functionals, i.e. PW91\cite{PERDEW:1992p6521} and RPBE\cite{Hammer:1999p9332}, and both failed to predict a minimum as well.
As for the other two atomic species, LDA gives binding for Ag on graphene, with the ordering top-bridge-hollow, see Fig.  \ref{fgr:AgLDA}, but at least 11\%\, too short equilibrium distances.

\begin{table}[htdp]
\caption{Adsorption energies (eV) of Ag on graphene, calculated by four different approximations to the exchange-correlation functional, initial vertical Ag-graphene distance $h$ and vertical equilibrium distances h$_{\mathrm{equ}}$ (\AA), at three binding sites, top (t), bridge (b), and hollow (h).}
\begin{center}
\begin{tabular}{|c|cc|cc|cc|}
\hline
& \multicolumn{2}{c|} {top} & \multicolumn{2}{c|} {bridge} & \multicolumn{2}{c|} {hollow}\\
& $E_{\mathrm{ads}}^{t}$ & $h_{\mathrm{equ}}^{t}$ & $E_{\mathrm{ads}}^{b}$ &  $h_{\mathrm{equ}}^{b}$ & $E_{\mathrm{ads}}^{h}$ & $h_{\mathrm{equ}}^{h}$  \\ \hline
PBE+D2& \quad -0.703 & 2.90 & \quad -0.700 & 2.90 & \quad -0.711 & 2.90 \\ 
vdW-DF & \quad -0.195 & 3.30 & \quad -0.194 & 3.40 & \quad -0.192 & 3.40\\
PBE & \multicolumn{2}{c|} {no binding} & \multicolumn{2}{c|} {no binding} & \multicolumn{2}{c|} {no binding} \\ 
LDA &  \quad -0.368 & 2.34 & \quad -0.359 & 2.35 & \quad -0.296 & 2.59 \\ \hline
\end{tabular}
\end{center}
\label{tableAg}
\end{table}%

In contrast to Cu and Au on graphene, where one finds a small hybridization in the density of states, i.e. a chemical contribution to the binding, PBE+D2 and vdW-DF give a pure physisorption of Ag on graphene.\cite{Amft:2010b}
However, the energies of the preferred binding sites are ordered differently in the two methods.
The minima in the total energy in PBE+D2 are ordered as hollow - top - bridge, while vdW-DF gives top - bridge - hollow.

The differences in the total energies, listed in Table \ref{tableAgdiff}, of the three binding sites are smaller than 11 meV in both vdW methods.  
In combination with the large binding distances,  see Table \ref{tableAg}, we conclude that both methods predict pure physisorption.
Also from Table \ref{tableAgdiff} one can conclude that Ag is the most mobile of the three coinage metals on graphene and does not have preferred diffusion paths as the other two.

The distortion of the graphene sheet by Ag is negligible for the equilibrium distances, see lower panels in Figs \ref{fgr:AgD2} and \ref{fgr:AgvdW}, supporting the interpretation that it is pure physisorption in this system.
The buckling away from the adatom at smaller distances is again a steric effect, as in the other two cases.

Our PBE+D2 results for Ag on graphene can be compared to the results in Ref. [\onlinecite{Jalkanen:2007p10951}], where van der Waals interactions also have been included according to the Grimme scheme. \cite{Grimme:06}
In contrast to our PBE+D2 results in Table \ref{tableAg}, Ref. [\onlinecite{Jalkanen:2007p10951}] found smaller adsorption energies, i.e. $-0.42$ to $-0.56$ eV, and a different binding order.
These calculations predicted the top site being more favorable than the bridge and hollow site, in accordance with our vdW-DF results.
Two methodological differences, which might explain the contradicting results between our PBE+D2 results and those of Ref. [\onlinecite{Jalkanen:2007p10951}], need to be pointed out.
First, the value of the $s_{6}$ coefficient, cf. Equ. \ref{e4}, is not clear from Ref. [\onlinecite{Jalkanen:2007p10951}].
Second, the Grimme method \cite{Grimme:06} is parametrized for PBE and not for PW91, which has been used in Ref. [\onlinecite{Jalkanen:2007p10951}] for the initial relaxation.

\begin{table}[htdp]
\caption{Since Ag is purely physisorpt on graphene the difference in the total energies $\Delta E_{0}$ (meV) at the three adsorption sites are of little significance.}
\begin{center}
\begin{tabular}{|c|c|c|c|}
\hline
& top to bridge & top to hollow & bridge to hollow \\ 
& $\Delta E_{0}^{t-b}$ & $\Delta E_{0}^{t-h}$ & $\Delta E_{0}^{b-h}$ \\ \hline  
PBE+D2 & 2 & 8 & 11  \\ 
vdW-DF & 1 & 3  & 2  \\ 
PBE & --- & --- & ---   \\
LDA & 9 & 72 & 62 \\ \hline
\end{tabular}
\end{center}
\label{tableAgdiff}
\end{table}%

\subsection{Au on graphene}
\begin{figure}[hbt]
\includegraphics[width=0.95\linewidth]{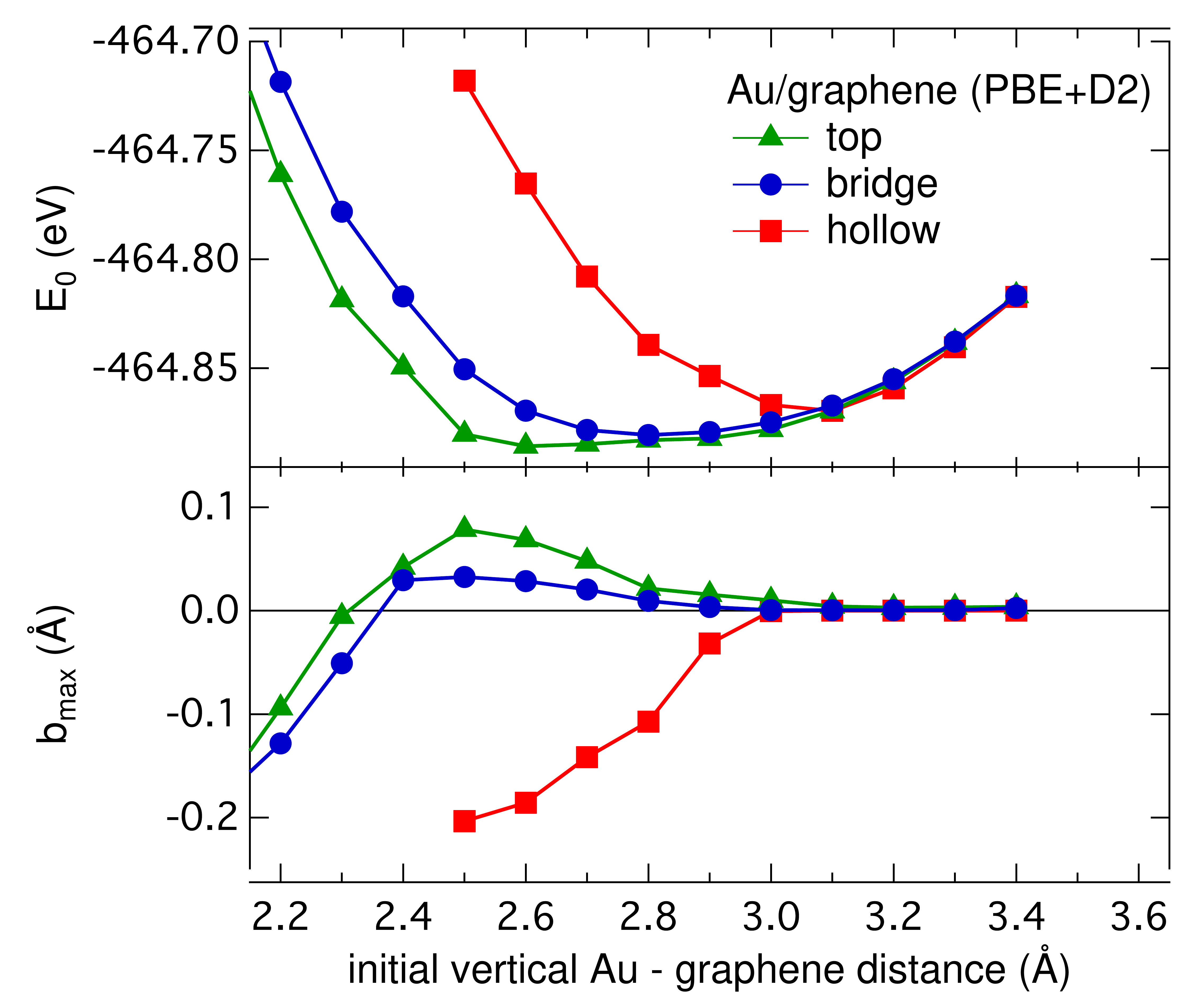}  
 \caption{(Color online) Upper panel: total energy as a function of the initial vertical Au adatom - graphene sheet distance $h$ for three binding sites, calculated in PBE+D2. Lower panel: vertical distortion ($b_{\mathrm{max}}$) of the carbon atoms in the first coordination shell of the Au adsorption site. Positive values of $b_{\mathrm{max}}$ mean the carbon atoms buckle towards the adsorbed adatom.}
\label{fgr:AuD2}
\end{figure}

\begin{figure}[hbt]
\includegraphics[width=0.95\linewidth]{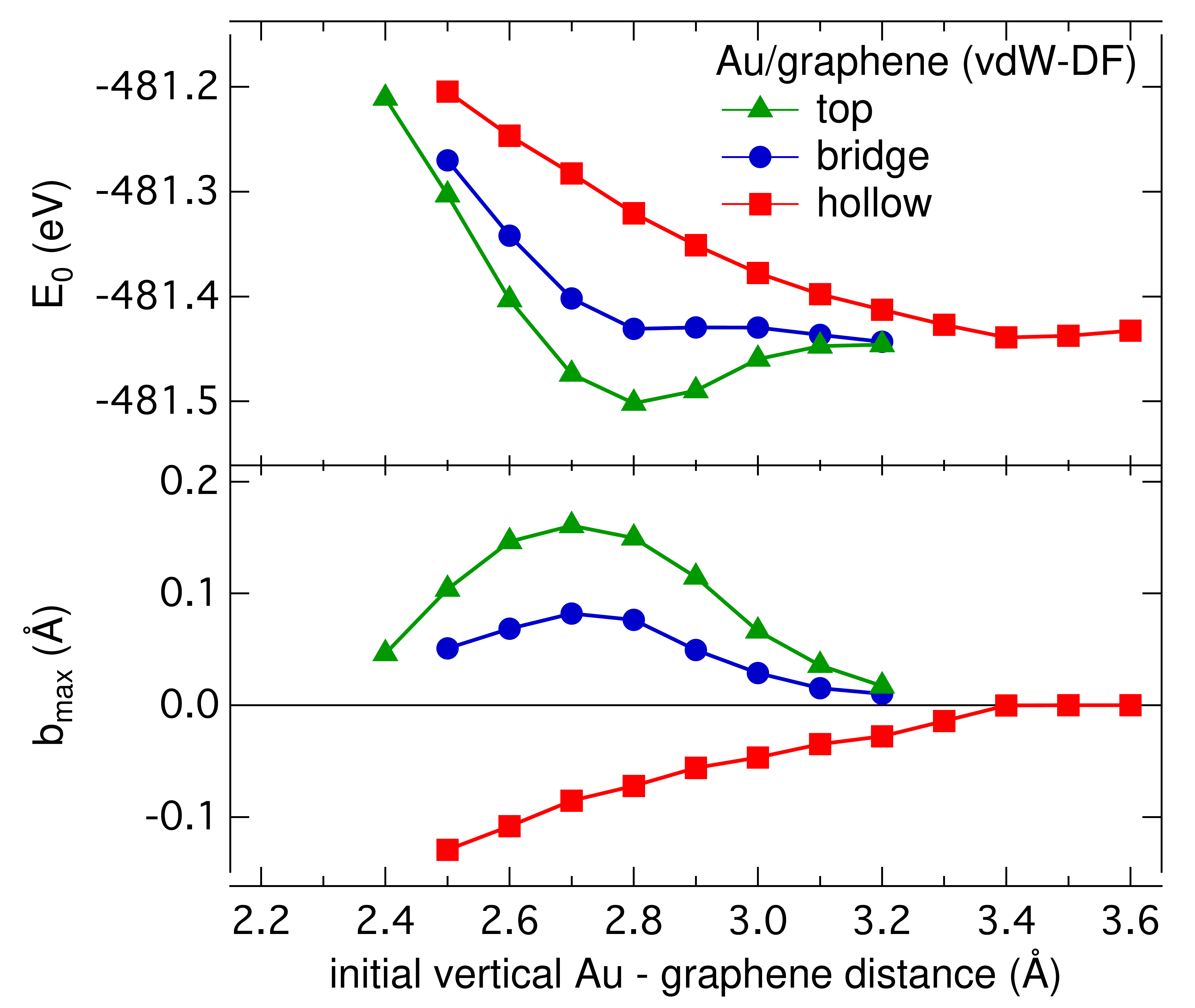}  
 \caption{(Color online) Upper panel: total energy   vs. initial vertical Au  - graphene distance for three binding sites, calculated in vdW-DF. Lower panel: vertical distortion ($b_{\mathrm{max}}$) of the carbon atoms closest to the Au adsorption site.}
\label{fgr:AuvdW}
\end{figure}

In the last studied case of a gold adatom on graphene all four studied approximations to the exchange-correlation energy predict the top adsorption site to be most favorable, see upper panels in Figs \ref{fgr:AuD2} and \ref{fgr:AuvdW}, as well as in Figs \ref{fgr:AuPBE} and \ref{fgr:AuLDA} for comparison.
Note that for the bridge site vdW-DF predicts a local minimum and PBE an inflection point at 2.8\,\AA\, only.
For the hollow site vdW-DF gives a shallow minimum at 3.4\,\AA, which is in fact lower in energy than the local minimum at the bridge site, while PBE fails to predict binding at this site.
Although, the total energy curve of the PBE+D2 method has a minimum at the hollow site as well, the equilibrium distance is 3.1\,\AA\, i.e. even larger than in the case of Ag on silver, which indicates  physisorption at this site, too.
Comparing the binding energies of PBE and PBE+D2, the latter essentially adds $-0.8$ eV to $E_{\mathrm{ads}}$ of the former.
This indicates the importance of the vdW interaction to the binding, which is even more pronounced in this system than for Cu on graphene.

\begin{table}[htdp]
\caption{Adsorption energies (eV) of Au on graphene, calculated by four different approximations to the exchange-correlation functional, initial vertical Au-graphene distance $h$ and vertical equilibrium distances $h_{\mathrm{equ}}$ (\AA), at three binding sites, top (t), bridge (b), and hollow (h).  Note that vdW-DF and PBE do not predict binding of Au at the hollow site.}
\begin{center}
\begin{tabular}{|c|cc|cc|cc|}
\hline
& \multicolumn{2}{c|} {top} & \multicolumn{2}{c|} {bridge} & \multicolumn{2}{c|} {hollow}\\
& $E_{\mathrm{ads}}^{t}$ & $h_{\mathrm{equ}}^{t}$ & $E_{\mathrm{ads}}^{b}$ &  $h_{\mathrm{equ}}^{b}$ & $E_{\mathrm{ads}}^{h}$ & $h_{\mathrm{equ}}^{h}$  \\ \hline
PBE+D2&\quad -0.886 & 2.53 & \quad -0.881 & 2.79 & \quad -0.870  & 3.10 \\ 
vdW-DF & \quad -0.385 & 2.65 & \quad -0.314 & 2.72 & \quad -0.322 & 3.40\\ 
PBE &  \quad -0.099 & 2.54 & \quad -0.081 & 2.72 & \multicolumn{2}{c|} {no binding} \\
LDA &  \quad -0.732 & 2.22 & \quad -0.698 & 2.26 & \quad -0.451 & 2.40 \\ 
\hline
\end{tabular}
\end{center}
\label{tableAu}
\end{table}

For the calculated adsorption energies, Table \ref{tableAu}, we obtained trends that are similar to those of Cu on graphene, cf. Table \ref{tableCu}.
The PBE+D2 method predicts $E_{\mathrm{ads}}$ ranging from -0.870 to -0.886 eV, which exceed even the LDA results for this system.
PBE, on the other hand, predicts significantly lower  $E_{\mathrm{ads}}$ of less than -0.099 eV, only, while vdW-DF gives adsorption energies that lie in between those of PBE and PBE+D2, ranging from $-0.310$ to $-0.385$ eV.

The equilibrium distances $h_{\mathrm{equ}}$ shown in Table \ref{tableAu} agree rather well for PBE+D2 and vdW-DF, i.e. with deviations of less than 9\%.
Note that the buckling of the graphene sheet is 50\% smaller in PBE+D2 than in vdW-DF, i.e. PBE.
Even PBE gives comparable $h_{\mathrm{equ}}$ for the top and bridge site, while LDA overbinds and predicts $h_{\mathrm{equ}}$ that are up to 23\%\, shorter than those obtained in PBE+D2.

\begin{table}[htdp]
\caption{Identifying the diffusion path of Au on graphene from the difference in the total energies $\Delta E_{0}$ (meV) at the three adsorption sites.}
\begin{center}
\begin{tabular}{|c|c|c|c|}
\hline
& top to bridge & top to hollow & bridge to hollow \\ 
& $\Delta E_{0}^{t-b}$ & $\Delta E_{0}^{t-h}$ & $\Delta E_{0}^{b-h}$ \\ \hline  
PBE+D2 & 5 & 16 & 11  \\ 
vdW-DF & 71 & 63 & 8 \\
PBE & 18 & --- & ---   \\ 
LDA & 34 & 281 & 247 \\  \hline
\end{tabular}
\end{center}
\label{tableAudiff}
\end{table}%

As in the case of copper, the differences of the total energies between the top and bridge site vary greatly among  the employed vdW approximations.
Still, one can conclude that the gold adatom is likely to diffuse along the carbon-carbon bonds.
Therefore, the usage of PBE is justified for studies of, for instance, the mobility and clustering of gold on graphene as in Ref. [\onlinecite{Amft:2010b}].

\section{Summary and conclusions}

We performed a systematic DFT investigation of the adsorption of copper, silver, and gold adatoms on graphene, especially taking van der Waals interactions by the vdW-DF and the DFT+D2 methods into account. 
For copper and gold we found that the PBE parametrization to the exchange-correlation energy predicts the same ordering of the adsorption sites as vdW-DF and PBE+D2, i.e. it also favors the top over the bridge and hollow positions.
We also find that the non-local interactions increase the calculated adsorption energies of Cu and Au on graphene by up to 0.8 eV (PBE+D2) and 0.45 eV (vdW-DF), respectively.

The predicted vertical equilibrium distance calculated with PBE, for Cu and Au adsorption, agrees to better than 13\%\, with the more advanced non-local methods.
Taking vdW interactions into account during the structural relaxations in the PBE+D2 method did not significantly change the buckling of the graphene sheet in the case of Cu and Ag compared to PBE calculations. 
Compared to the PBE calculation, we found a 50\% smaller buckling of the graphene sheet in the case of gold with the PBE+D2 method that might be related to its parametrization.
Ripples in the graphene structure give rise to both effective scalar and vector potentials in the electronic structure of graphene, see e.g Ref. [\onlinecite{Gibertini:2010p11422}]. 
The adsorption of Cu or Au atoms may be an important tool to enhance this rippling and hence the strength of these scalar and vector potentials.
For the adsorption of silver on graphene our calculations suggest that it is purely of van der Waals type.
Different generalized gradient  approximations to the exchange-correlation functional fail to give any binding at all, while the PBE+D2 and vdW-DF predict physisorption at large equilibrium distances of 2.9 and 3.3\,\AA, respectively.
Also the calculated distortion of the graphene sheet upon adsorption of a silver atom  was found to be negligible.
From the differences in the total energies we conclude that diffusion of Cu and Au takes place along the carbon-carbon bonds, while the Ag adatoms can diffuse almost unrestricted on the graphene sheet.

\begin{acknowledgments}
This research was supported by the Swedish Energy Agency (Energimyndigheten), the Swedish Research Council (Vetenskapsr\aa det), and Research and Innovation for Sustainable Growth (VINNOVA).
O. E. is grateful to the European Research Council (ERC) for support.
S. L. acknowledges financial support from ANR Grant ANR-07-BLAN-0272 and ANR Grant ANR-06-NANO-053-02. 
Supercomputer time was granted by the Swedish National Infrastructure for Computing (SNIC) and GENCI (project x2010085106).
\end{acknowledgments}

\section{Appendix}

In the appendix we present calculated total energy curves and vertical distortions of the carbon atoms closest to the adsorption sites for the PBE and LDA functional. 
The Cu results are found in section A, the Ag results in section B and the Au results in section C.

\subsection{Cu/graphene (PBE and LDA)} 

\newpage

\subsection{Ag/graphene (PBE and LDA)} 

\newpage
\subsection{Au/graphene (PBE and LDA)} 

\clearpage

\bibliography{references}

\end{document}